\documentclass[aps,twocolumn,showlabels,showrefs,amsmath,amssymb,prl,superscriptaddress,floatfix,colors]{revtex4}

\usepackage{lineno}
\usepackage{graphicx}
\usepackage{dcolumn}
\usepackage{bm}

\usepackage{graphicx}
\usepackage{dcolumn}
\usepackage{bm}
\usepackage{amssymb}
\usepackage{hyperref}
\usepackage{multirow}
\usepackage{color}
\usepackage[normalem]{ulem}

\usepackage[cp1251]{inputenc}

\makeatletter
\newcommand*{\sumcirclearrowleft}{%
 \DOTSB
 \mathop{
  \mathchoice
   {\rlap{\kern.25em\rotatebox[origin=c]{-90}{$\circlearrowleft$}}{\sum}}
   {\vcenter{\rlap{\kern.2em\rotatebox[origin=c]{-90}{$\scriptscriptstyle\circlearrowleft$}}}{\sum}}
   {\sum}{\sum}
 }\slimits@
}

\newcommand*{\sumcirclearrowright}{%
 \DOTSB
 \mathop{
  \mathchoice
   {\rlap{\kern.25em\rotatebox[origin=c]{90}{$\circlearrowright$}}{\sum}}
   {\vcenter{\rlap{\kern.2em\rotatebox[origin=c]{90}{$\scriptscriptstyle\circlearrowright$}}}{\sum}}
   {\sum}{\sum}
 }\slimits@
}
\makeatother

\begin{document}

\title{Defects Superdiffusion and Unbinding in a 2D XY Model of Self-Driven Rotors}


\author{Ylann Rouzaire}
\affiliation{Institute of Physics, \'Ecole Polytechnique F\'ed\'erale de Lausanne, 1015 Lausanne, Switzerland}
 \affiliation{Departament de F\'i­sica de la Materia Condensada, Universitat de Barcelona, Mart\'i­ i Franqu\`es 1, E08028 Barcelona, Spain}
\author{Demian Levis}%
 \affiliation{Departament de F\'i­sica de la Materia Condensada, Universitat de Barcelona, Mart\'i­ i Franqu\`es 1, E08028 Barcelona, Spain}
 \affiliation{UBICS University of Barcelona Institute of Complex Systems , Mart\'i­ i Franqu\`es 1, E08028 Barcelona, Spain}

\date{\today}

\begin{abstract}
We consider a non-equilibrium extension of the two-dimensional (2D) XY model, equivalent to the noisy Kuramoto model of synchronization with short-range coupling, where rotors sitting on a square lattice are self-driven by random intrinsic frequencies. We study the static and dynamic properties of topological defects (vortices) and establish how self-spinning affects the Berezenskii-Kosterlitz-Thouless phase transition scenario. The non-equilibrium drive breaks the quasi-long-range ordered phase of the 2D XY model into a mosaic of ordered domains of controllable size and results in self-propelled vortices that generically unbind at any temperature, featuring superdiffusion $\langle r^2(t)\rangle\sim t^{3/2}$ with a Gaussian distribution of displacements. Our work provides a simple framework to investigate topological defects in active matter and sheds new light on the  problem of synchronization of locally coupled oscillators. 
\end{abstract}

\maketitle

In equilibrium, the competition between noise and interactions is responsible for the different phases of matter and the transitions between them. The situation is particularly interesting at low dimensions, where certain collective coherent structures, topological defects, drive a Berezenskii-Kosterlitz-Thouless (BKT) phase transition between a disordered phase and a phase with quasi-long-range order (QLRO) \cite{Berezinskii1971,Kosterlitz1973, Kosterlitz1974}. Although such scenario has been originally established from the study of the two-dimensional (2D) XY model, a BKT transition generically takes place in 2D systems as soon as topological defects interact logarithmically, thus applying to diverse contexts such as boson gases \cite{DalibardBKT}, superconductors \cite{Hebard1980}, melting \cite{Halperin1978}, liquid crystals \cite{SinghRev}, among others. 
Out-of-equilibrium, the situation is far less clear, as we lack a general guide to assess how matter self-organizes. Active matter made of self-driven constituents constitute a particularly interesting class of non-equilibrium systems
 \cite{MarchettiRev, RamaswamyRev}. In recent years, a great deal of attention has been paid to the study of phase transitions in 2D  (off-lattice) systems of self-propelled particles, 
and the study of defects has mainly been focused on active nematics \cite{SaguesRev, ZhangRev, Thampi2013,Giomi2013, Giomi2014, Decamp2015, ChateBacteria2019, Shankar2018, Shankar2019, Gompper2020}
or dense assemblies of spherical particles 
\cite{Weber2014, PaliwalDijkstra, LinoDefects, Bartolo2021, ChardacBartolo2021}. These studies reveal that topological defects in active systems generically self-propel. In particular, very recent experiments on monolayers of self-spinning colloids report the superdiffusion of dislocations with $\langle r^2(t)\rangle\sim t^{3/2}$ \cite{Bartolo2021}. 
Despite all these efforts, the question of how the general BKT scenario is affected by a local energy input, and how the resulting defects move and interact with each other, remains largely open.

To address this question, we go back to the earliest example of a topological phase transition,  and 
study a natural out-of-equilibrium extension of the 2D XY model, in which spins, or rotors, are self-driven by random intrinsic frequencies. In other words, we consider a Kuramoto model, a paradigm for synchronization,  with short-range coupling in the presence of noise \cite{KuramotoBook, AcebronRev, StrogatzBook, Pikovsky2003}.
Although in its original formulation \cite{KuramotoOriginal}, the model considered an all-to-all coupling, 
many extensions have been considered to describe different situations \cite{ArenasRev,RuffoRev}. However, studies of oscillators with short-range interactions are relatively scarce and exclusively focused on noiseless dynamics \cite{Strogatz1988, Sakaguchi1987, Daido1988, Niebur1991, Lee, Hong2005, Hong2007, Sarkar2021}, although including noise is both more realistic in general (in particular at the micro-scale) and renders the connection between synchronization and critical phenomena clearer. 
As such, our framework offers a platform to investigate fundamental questions of matter out-of-equilibrium, bridging together the study of topological defects in active matter and the synchronization of noisy oscillators.

\begin{figure}[h!]
\includegraphics[width=0.9\linewidth]{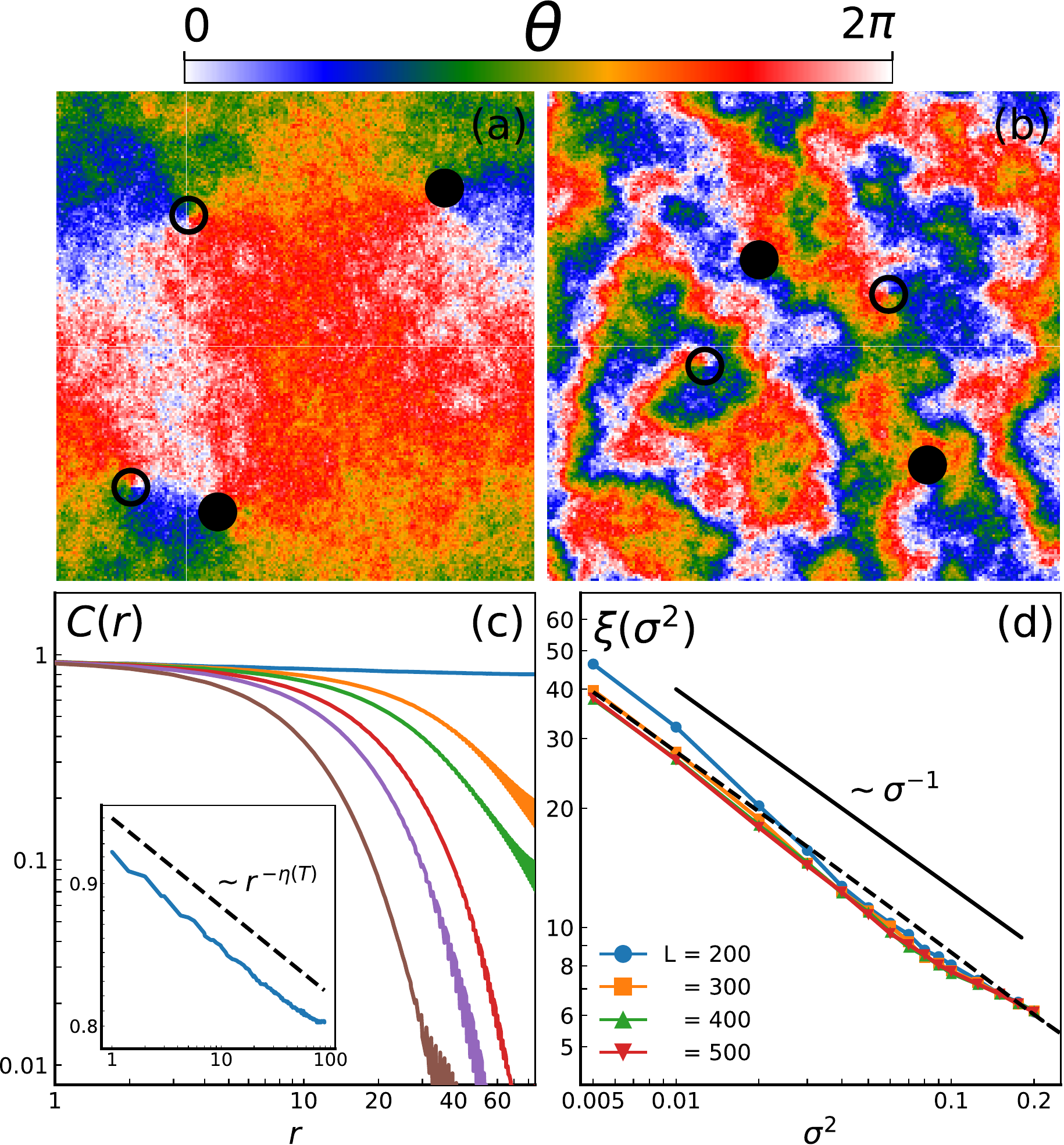}
\caption{
Snapshots for $T =0.2$, (a) $\sigma^2=0$ and (b) $\sigma^2=0.1$. The phase of each oscillator is represented by a color scale. Vortices (antivortices) are represented by filled (empty) circles.
(c) $C(r)$ (in log-log) at $T=0.2$ and $\sigma^2 = 0,0.005,0.01,0.02,0.03,0.06$ (from top to bottom). Inset: the $\sigma^2=0$ case, showing $C(r)\sim r^{-\eta}$, with $\eta(T)=T/2\pi$.
(d) $\xi$ vs. $\sigma^2$ for $T =0.2$ and different $L$. The dashed line represents the prediction $\ell{}={\pi}/({2\overline{|\Delta \theta_{0}|}})$ (see text).
}\label{fig:snap-corr}
\end{figure}

In this Letter, we establish the general collective behavior of the self-driven 2D XY model.
We show that, while in equilibrium vortices are diffusive and bind at low temperatures, in the presence of self-driving they unbind and become \emph{superdiffusive}. 
 As a result, the mechanism triggering a BKT transition in the XY model is lost. At low frequency dispersion $\sigma$ and temperature $T$, vortices are rare, yet QLRO is destroyed, resulting in a patchwork of finite-size ordered domains whose size scales as $1/\sigma$ (see Fig.~\ref{fig:snap-corr}), leaving behind smooth boundaries on which vortices surf. 

We consider a set of rotors (or spins) arranged on a $L\times L$ square lattice with periodic boundary conditions (PBC). The evolution of their phase $\theta_i$ follows 
\begin{equation} \label{eq:model} 
 \dot{\theta}_i=\Omega_i + J \sum_{ j\in\partial_i} \sin(\theta_j - \theta_i) + \sqrt{2D}\nu_i(t) \, .
 \end{equation}Here the sum runs over the nearest-neighbors of $i$. The intrinsic frequencies $\Omega_i$ are drawn from a zero-mean Gaussian distribution with variance $s^2$ and $\nu_i$ is a Gaussian white noise of unit variance. To understand the key control parameters, we express times in units of $J^{-1}>0$, leading to ${\omega}_i=\Omega_i/J$, $\sigma= s/J$ and $T=D/J$, the reduced temperature. We numerically integrated Eq.~(\ref{eq:model}) (see SM \cite{SM} for details), typically using $L=200$ (we went up to $L=500$ to explore finite-size effects).
 
For $\sigma=0$, Eq.~(\ref{eq:model}) arises from the XY Hamiltonian $H=-J\sum_{\langle i,j \rangle}\bold{S}_i\cdot\bold{S}_j$ and $\bold{S}_i=(\cos \theta_i,\sin\theta_i)$, with overdamped (non-conserved) dynamics \cite{LoftDeGrand,YurkeHuse1993,Paoluzzi2018}.
 Intrinsic frequencies drive the XY model out-of-equilibrium as they account for a constant energy injection at the level of each spin. 
 In equilibrium ($\sigma=0$), the system exhibits a BKT transition at $T_{KT}\approx 0.89$ \cite{Kosterlitz1973, Kosterlitz1974, Gupta1988, Hasenbusch2005}, characterized by a change in behavior of the correlation function $C(r) = \langle\bold{S}_i\cdot\bold{S}_j \rangle_{|\bold{r}_i-\bold{r}_j| = r}$ (the brackets denote the average over different realizations of noise and the intrinsic frequencies). Below $T_{KT}$ vortex-antivortex pairs bind, giving rise to critical correlations. 
 As illustrated in Fig.~\ref{fig:snap-corr}(a), in the low-$T$ regime, the system comprises large regions of spins sharing the same color, and one can identify  visually which vortices are bounded. On the contrary, upon self-spinning, the phase field disturbances induced by the vortices 
 do not extend over large distances (see Fig.~\ref{fig:snap-corr}(b)). Ordered regions are in this case localized, and one can no longer pair vortices by eye. 
The computation of $C(r)$ supports this picture: for $\sigma=0$ correlations in the low-$T$ phase decay algebraically $C(r)\sim r^{-\eta(T)}$, with $\eta(T)=T/2\pi$, while for $\sigma>0$, $C(r)$ decays exponentially, defining a correlation length $\xi$. 
 
We extract $\xi$ from $C(\xi) = 1/e$ and plot its evolution with $\sigma$ at $T=0.2$, in Fig.~\ref{fig:snap-corr}(d)
It displays a $\sigma^{-1}$ decay, which can be understood with the following 1D model. We denote the phase and intrinsic frequency difference between each pair of neighbors on a chain as $\Delta\theta_i=\theta_{i+1}-\theta_i$ and $\Delta\omega_i=\omega_{i+1}-\omega_i$. 
The phase difference accumulated along $n$ links reads $\delta_n=\sum_{j=i}^{i+n}|\Delta\theta_j|$. 
We consider that two spins at a distance $\ell{}$ are uncorrelated when $\delta_{\ell}=\pi/2$, allowing us to define a characteristic length (see \cite{SM} for details).
We then assume that $\Delta\theta_i$ is given by $\Delta \theta = \sin^{-1}(\Delta\omega/2)$, the steady value for a single pair of oscillators, and thus $\sum_{j=i}^{i+\ell{}}|\Delta\theta_j|\approx \ell\overline{|\Delta \theta|}$, where the overline denotes its mean, computed from the probability distribution
$f(\Delta \theta) = \frac{\cos\, \Delta \theta}{\sigma\,\sqrt{\pi}}\,\exp\left\{-\left(\frac{\sin\, \Delta \theta}{\sigma}\right)^2\right\}$. 
As shown in Fig.~\ref{fig:snap-corr}(d), the results obtained from $\ell{}=\frac{\pi}{2}{\overline{|\Delta \theta|}}$ reproduce the numerical values of $\xi$. 

\begin{figure}[b]
\includegraphics[width=\linewidth]{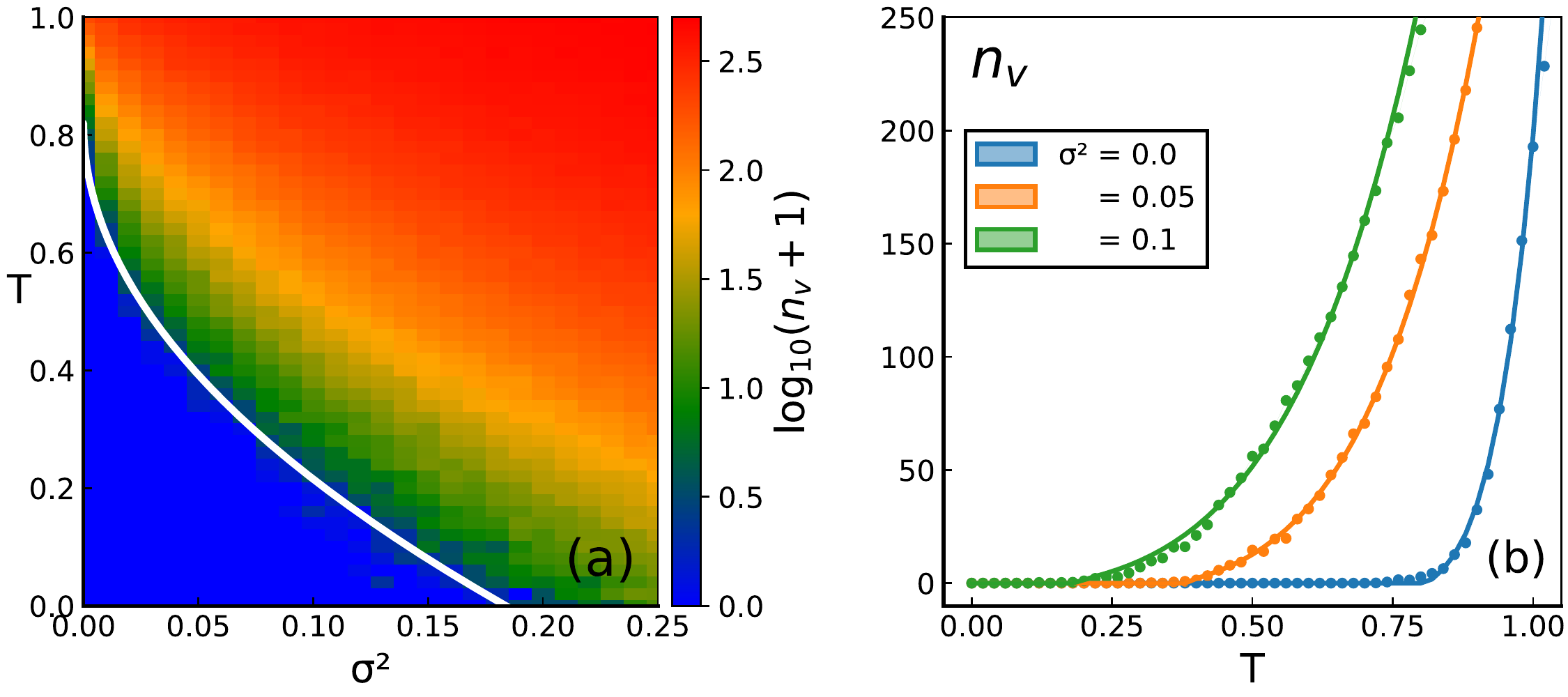} 
\caption{ (a) Number of free vortices $n_v$ + 1 represented on a logarithmic color scale, in the $\sigma^2-T$ plane. 
The white line shows Eq.~(\ref{eq:Tc}) with $a=0.43$.
(b) $n_v$ as a function of $T$ for $\sigma^2=0,\, 0.05,\, 0.1$. Solid lines show a two-parameter fit to the BKT form $n_v=\exp(\alpha\sqrt{(T-T_c)/T_c})$ for $T>T_c$, with $T_c=0.82$, $0.39$, $0.20$ and $\alpha=11.2$, $4.8$, $3.2$ for $\sigma^2=0,\, 0.05,\, 0.1$, respectively.
}\label{fig:rhod}
\end{figure}

The previous analysis shows that self-spinning destabilizes QLRO, generating ordered domains of finite and controllable size $\propto\sigma^{-1}$. 
In the XY model, QLRO is lost due to the unbinding of vortices: the high-$T$ phase is rich in vortices, interpreted in the Coulomb gas picture, as a conducting phase made of free diffusive charges. 
We thus turn our attention into vortices. 
A vortex (antivortex) is identified as a square plaquette with a winding number $q=\pm1$, defined by the sum of the phase differences along its four bonds $\sumcirclearrowleft \Delta \theta_{i,j}=2\pi q$. To distinguish free from bounded vortices,
we first pair vortices and anti-vortices so that the sum of the pair separations is minimized. Once the optimal matching has been found (using the so-called Hungarian algorithm \cite{Hungarian, HungarianComplexity, HungarianImproved}), a pair of vortices are said to be free if distant of more than $h=\max(3,L/N_p)$, $N_p$ being the number of vortex pairs in the system (see \cite{SM} for details). The number of free vortices $n_v$ thus obtained is shown in Fig.~\ref{fig:rhod} (a) as a function of both $T$ and $\sigma$. Below a critical temperature $T_c(\sigma)$, $n_v\!\approx\!0$. Above $T_c(\sigma)$, the number of free vortices grows exponentially, as expected through a BKT transition, see Fig.~\ref{fig:rhod} (b). 
However, below $T_c$, spatial correlations are also short-ranged: there is no BKT transition for $\sigma>0$. A finite-size analysis (see \cite{SM}) reveals that 
 \begin{equation}
	T_c (\sigma)= T_{KT}\left( 1-\,a\,\sigma\,{\ln L}\right) \ ,
  \label{eq:Tc}
\end{equation}
which in the $T=0$ limit is consistent with \cite{Lee}. 
Note that the $n_v\!=\!0\,$-region shrinks continuously as $L$ increases, going to (logarithmically) the XY critical point in the thermodynamic limit.
 
 \begin{figure}[b]
\includegraphics[width=\linewidth]{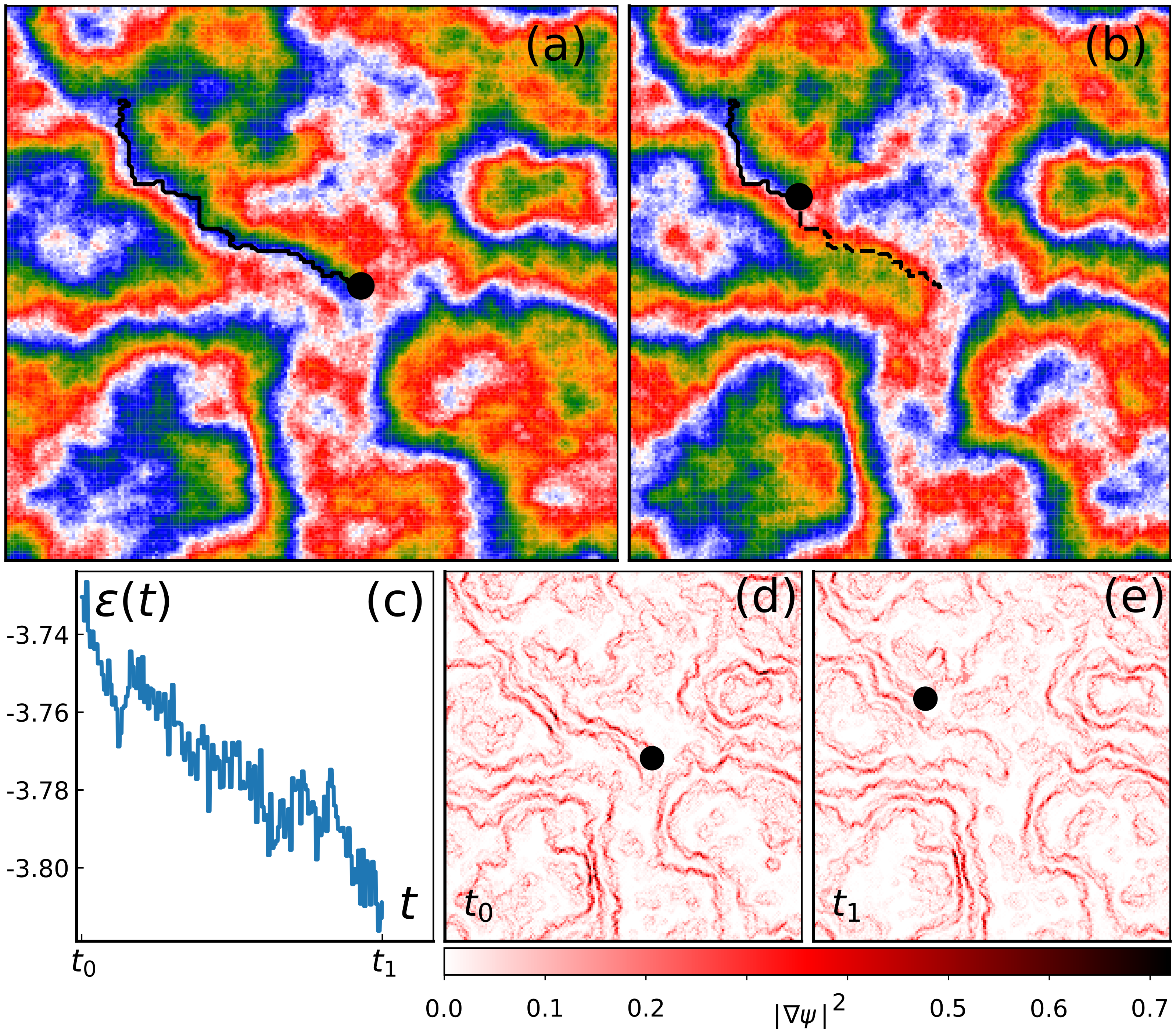}
\caption{Typical vortex trajectory at $T = 0.05, \sigma^2=0.1$. (a) Snapshot of the system at $t_0$, showing a vortex with a dot alongside a section of its (upcoming) trajectory and (b) at $t_1=t_0+500$, showing the path followed by the vortex in broken lines  (same color code as Fig.~\ref{fig:snap-corr}). 
(c) Decay of the energy per spin of an area around the domain boundary due to the motion of the vortex. Panels (d) and (e) display $|\nabla \psi|^2$ at $t_0$ and $t_1$, showing how the vortex removes the boundary it had just ridden. 
}\label{fig:vortexmotion}
\end{figure}

The mechanism responsible for the absence of QLRO is not due to the creation of defects but rather to their dynamics. 
Self-driving sets finite-size domains, generating a network of (smooth) boundaries which store most elastic energy in the system.
Such energy imbalance across the boundaries favors the presence of vortices along them. 
As shown in Fig.~\ref{fig:vortexmotion} (and the movies \cite{SM}), vortices move preferentially along domain boundaries and relieve excess elastic energy by erasing them in their wake, resulting in a ballistic-like motion, akin self-propelled particles. To illustrate this picture, we also show in Fig.~\ref{fig:vortexmotion} maps of $|\nabla{\psi}|^2$, where the local (scalar) order parameter $\psi (\bold{r})$ is defined as the projection of the spin field onto the mean orientation of the system.
 The structure of the system is quite dynamic: domains are perpetually reshaping and emerging. In turn, vortices' fate is to move along the resulting time-evolving network of paths. 
 
\begin{figure}[h]
\includegraphics[width=0.98\linewidth]{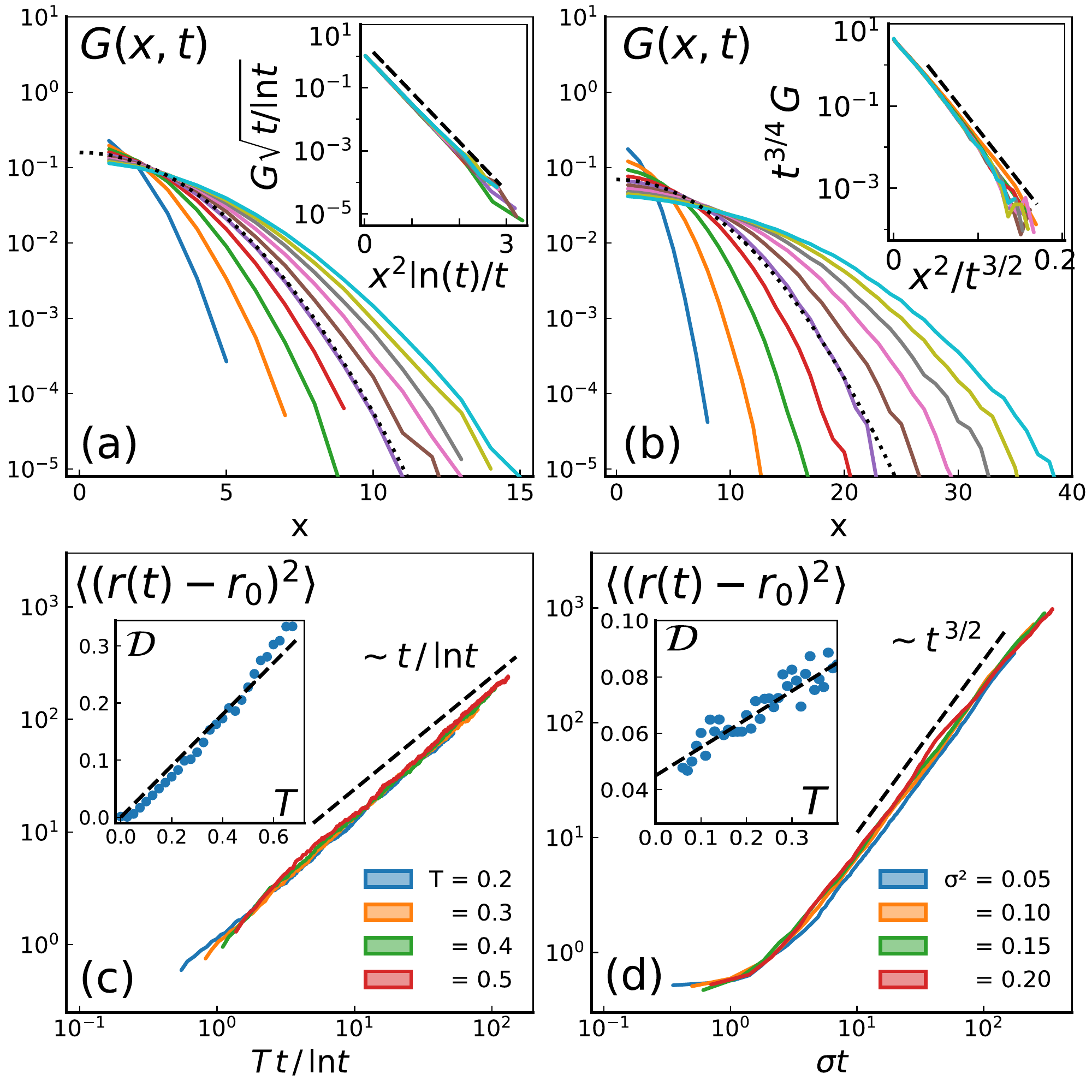}
\caption{\label{fig:gmsd} (a) Distribution of vortex displacements $G(x,t)$ at different times $t=50,100,...,500$ (from left to right) and fixed $\sigma=0$, $T=0.2$. The dotted line shows Eq.~(\ref{eq:GXY}) with $\mathcal{D}=0.069$ and $t=250$.
Inset: these curves collapse when using the scaling Eq.~(\ref{eq:GXY}). The exponential decay of slope $1/(4\mathcal{D})$ is shown in dash. 
(b) $G(x,t)$ (same times as in (a)) for $\sigma^2=0.025$ and $T=0.2$. The dotted line shows Eq.~(\ref{eq:GNKM}) with $\mathcal{D}=0.066$ and $t=250$. The inset shows an exponential decay of slope $1/(4\mathcal{D}\sigma^{3/2})$. 
(c) MSD of a single vortex  for $\sigma=0$ and different $T$. 
The dashed line shows a logarithmic correction to normal diffusion. Inset: $\mathcal{D}$ as a function fo $T$, confronted to a linear growth of slope $0.45$. 
(d) MSD at fixed $T=0.05$ but different $\sigma^2$, as a function of $\sigma t$. The dashed line shows  $\sim t^{3/2}$. Inset: $\mathcal{D}$ vs. $T$ for $\sigma^2=0.025$, confronted to a linear growth of slope $0.1$. 
}
\end{figure}

We investigate the dynamics of vortices by tracking them in our simulations. We prepare an initial configuration with a net winding number $q=1$, hosting a single vortex at the center of our simulation box. We then let the system evolve at low-$T$ and low-$\sigma$, ensuring that (i) new defects are not spontaneously created and (ii) the simulation box is large enough such that boundary effects are negligible \cite{LoftDeGrand, Olsson1992, Olsson1999, Japo}. The probability distribution function (PDF) of displacements thus obtained, $G(x,t)$, is shown in Fig.~\ref{fig:gmsd} for different times. For the 2D XY model, it follows 
\begin{equation}\label{eq:GXY}
G(x,t)=\frac{1}{\sqrt{{4\pi \mathcal{D}} \,t/\ln t }}\exp \left[-\frac{x^2}{4\mathcal{D}\, t/\ln t}\right] \ .
\end{equation}
Such form implies anomalous diffusion, in the sense that the ensemble averaged mean square displacement (MSD) behaves, at long times, as
\begin{equation}\label{eq:MSDXY}
\langle\left[ \bold{r}(t)-\bold{r}(0)\right]^2\rangle=4\mathcal{D}\,t/\ln t\, ,
\end{equation}
showing the expected correction to the normal diffusion scaling, due to the logarithmic dependence of the defect mobility on its size, with a pseudo-diffusion coefficient $\mathcal{D}\propto T$ (see Fig.~\ref{fig:gmsd}(c)) \cite{YurkeHuse1993, Japo}. 

 Unlike passive diffusion in equilibrium conditions, upon self-spinning, vortices become super-diffusive, following a Gaussian PDF
\begin{equation}\label{eq:GNKM}
G(x,t)=\frac{1}{\sqrt{{4\pi \mathcal{D}} \,(\sigma t)^{3/2} }}\exp \left[-\frac{x^2}{4\mathcal{D}\, (\sigma t)^{3/2}}\right] \ ,
\end{equation}
which naturally results in 
\begin{equation}\label{eq:MSDNKM}
\langle\left[ \bold{r}(t)-\bold{r}(0)\right]^2\rangle=4\mathcal{D}(\sigma t)^{3/2} \ .
\end{equation}
As shown in Fig.~\ref{fig:gmsd}, these expressions describe our numerical data accurately. The $\sigma$-dependence of the vortex dynamics enters through a rescaling of time: the typical time associated with its motion along a domain is $\propto \xi\sim 1/\sigma$. 
All the $T$-dependance is simply contained in $\mathcal{D}\propto T$. Note that in the Kuramoto limit $T\to0$, vortices are still moving in a lively heterogeneous medium and are thus superdiffusive. 

Such behavior is reminiscent of the anomalous diffusion of tracer particles in quenched random velocity fields \cite{Redner1989superdiff, Redner1990superdiff, BouchaudGeorgesPRL1990, PhysA2020} or turbulent flows \cite{BouchaudRev}. However, superdiffusion in this classical context is typically characterized by non-Gaussian statistics and driven by convection rather than an internal drive. 
Recent studies in 2D systems of self-driven units have shown that the dynamics of topological defects is strongly affected by local energy inputs, resulting in spontaneous self-propulsion \cite{Thampi2013,Giomi2013, Giomi2014, Decamp2015, ChateBacteria2019, Shankar2018,Shankar2019, Gompper2020, Bartolo2021}. 
Most of these studies concern active nematics \cite{SaguesRev}, typically described at the coarse-grained level of hydrodynamic theories. Refs. \cite{Decamp2015,ChateBacteria2019, Gompper2020} constitute salient exceptions where active nematics are described at the level of their constituents and the motion of defects monitored.
 In particular, \cite{Gompper2020} reports superdiffusion of $+\frac{1}{2}$ disclinations, with $\langle \bold{r}^2(t)\rangle\sim t^{1.8}$, in dense systems of flexible active filaments populated, 
 and in \cite{Bartolo2021} dislocations in dense 2D systems of self-driven colloids show supperdiffusion with   $\langle \bold{r}^2(t)\rangle\sim t^{3/2}$.
It is thus remarkable that in our simpler lattice model, we recover similar behavior for vortices and anti-vortices, which are isotropic defects. In our case, their preferential direction of motion is given by the underlying domain structure.  
Moreover, we show that defects obey the PDF Eq.~(\ref{eq:GNKM}).

\begin{figure}[b]
\includegraphics[width=\linewidth]{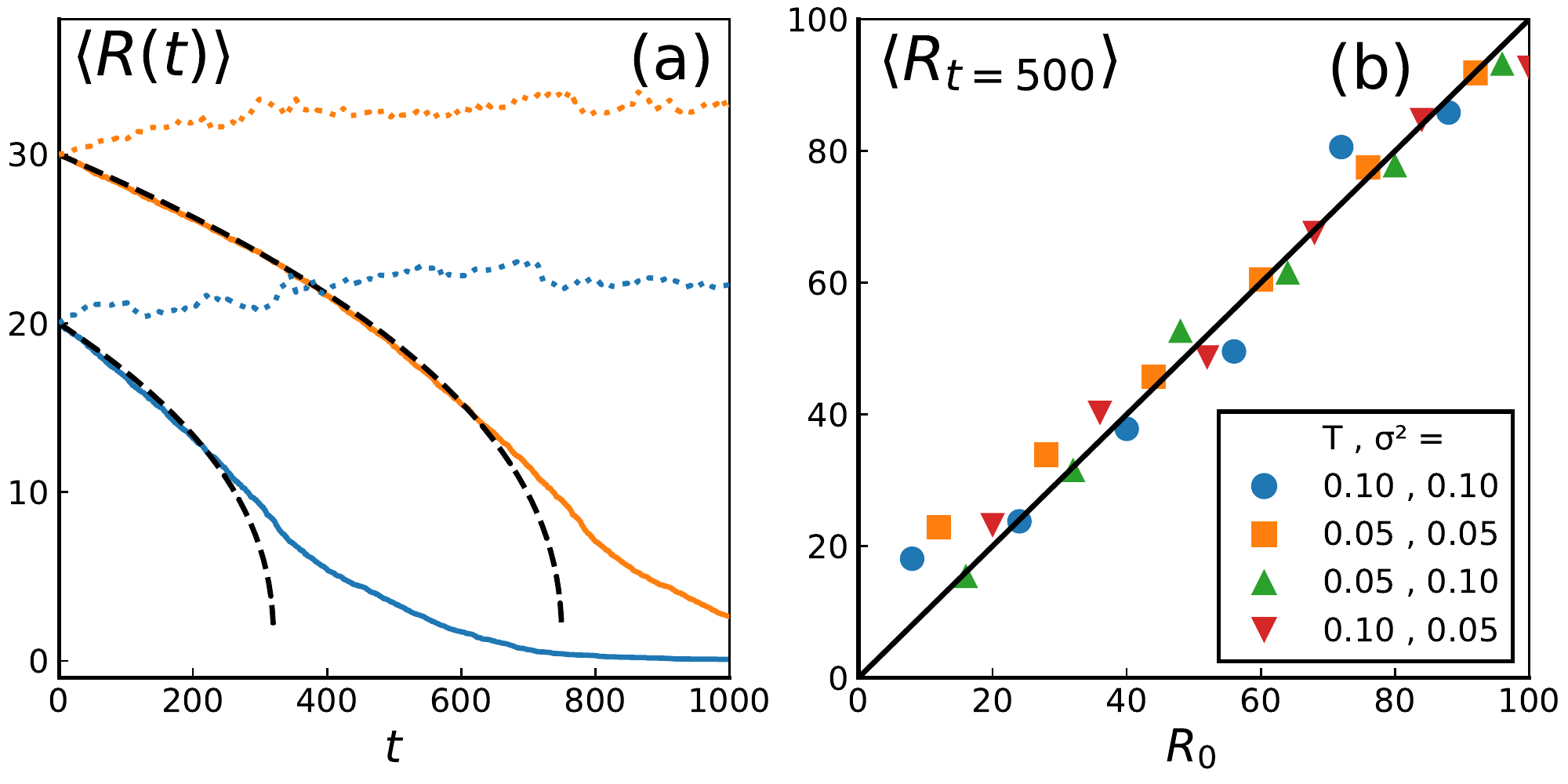}
\caption{ \label{fig:2vortices} (a) Vortex-antivortex separation for two different initial values $R_0$, averaged over 120 independent runs. Colored solid lines are trajectories for equilibrium vortices at $T=0.1$. Black dashed lines are the predictions for the 2D XY model \cite{YurkeHuse1993}. Broken lines correspond to the driven system at $T=0.1$ and $\sigma^2=0.1$. 
(b) Averaged pair separation after a time $t=500$, as a function of their initial separation for several values of $T$ and $\sigma$. The black line corresponds to $\langle R_{t=500}\rangle=R_0$. 
}
\end{figure}

In equilibrium, a BKT transition generically occurs as soon as topological defects interact logarithmically. We  investigate here the effective interaction between vortices by initializing the system with a vortex  anti-vortex pair at a distance $R_0$, enforcing PBC \cite{SM}.
We then let the system evolve and monitor the separation distance $R(t)$, averaged over different realizations of the dynamics, taking different initial separations and ensuring that the system hosts a single vortex-antivortex pair at all times. 
As confirmed in Fig.~\ref{fig:2vortices}(a), in the 2D XY model, a vortex-antivortex pair annihilates (see movies \cite{SM}). 
Moreover, its dynamics is consistent with the overdamped description $\ln(R)\dot{R} = -V'(R)$ in \cite{YurkeHuse1993}, where $V(R)=\ln R$.
 (The departure from the numerical data at small distances is explained in \cite{SM}). 

Self-spinning qualitatively alters this scenario. Vortices that would otherwise annihilate, no longer do. The analysis of pair trajectories (Fig.~\ref{fig:2vortices}) reveals that, in some cases, vortex pairs annihilate when bumping into each other by chance, but, in many other (most) cases, they move apart and never annihilate in the observation time window. As a result, their average distance remains largely constant over time, indicating that defects unbind and evolve freely. As shown in Fig.~\ref{fig:2vortices}(b), their separation remains, on average, identical to the initial one, in contrast with the behavior recently reported in dense monolayers of self-spinning colloids, where a threshold driving is needed in order to unbind defects \cite{Bartolo2021}. 

Our study provides a simple and general framework to understand the role of self-driving in 2D systems that would otherwise exhibit a BKT phase transition. 
The usual vortices of the 2D XY model unbind and behave as self-propelled particles, showing superdiffusion with Gaussian statistics, when spins are self-driven or active. 
The entanglement between the dynamics of vortices and the underlying spin structure is a consequence of activity. The formation of finite size structures in 2D active  systems has been reported in various cases \cite{GiomiPierce, SaguesRev, ClaudioPRL, Bartolo2021}, however, its connection with the BKT scenario in a lattice spin model has remained elusive. As such, our work provides a simple platform to investigate fundamental aspects of topological defects in active systems on general grounds.
Moreover, our results should be insightful for numerous synchronization studies, for instance, in monolayers of self-spinning colloids \cite{CicutaSyncRev,zhang2021quincke}, genetic clocks \cite{Mondragon2011,Prindle2014} or spin torque oscillators \cite{Ferran2016}, and  hope they will revigorate the study of coupled oscillators from the perspective of topological defects.


\paragraph{Acknowledgments}
We warmly thank Elisabeth Agoritsas and Matteo Paoluzzi for a critical reading of the manuscript. 
YR acknowledges the Swiss European Mobility Program (SEMP) for partial financial support.
DL acknowledges Ministerio de Ciencia, Innovaci\'on y Universidades MCIU/AEI/FEDER for financial support under grant agreement RTI2018-099032-J-I00.

\bibliography{biblio}

\providecommand{\noopsort}[1]{}\providecommand{\singleletter}[1]{#1}%
\begin{thebibliography}{63}
\expandafter\ifx\csname natexlab\endcsname\relax\def\natexlab#1{#1}\fi
\expandafter\ifx\csname bibnamefont\endcsname\relax
  \def\bibnamefont#1{#1}\fi
\expandafter\ifx\csname bibfnamefont\endcsname\relax
  \def\bibfnamefont#1{#1}\fi
\expandafter\ifx\csname citenamefont\endcsname\relax
  \def\citenamefont#1{#1}\fi
\expandafter\ifx\csname url\endcsname\relax
  \def\url#1{\texttt{#1}}\fi
\expandafter\ifx\csname urlprefix\endcsname\relax\def\urlprefix{URL }\fi
\providecommand{\bibinfo}[2]{#2}
\providecommand{\eprint}[2][]{\url{#2}}

\bibitem[{\citenamefont{Berezinskii}(1971)}]{Berezinskii1971}
\bibinfo{author}{\bibfnamefont{V.}~\bibnamefont{Berezinskii}},
  \bibinfo{journal}{Sov. Phys. JETP} \textbf{\bibinfo{volume}{32}},
  \bibinfo{pages}{493} (\bibinfo{year}{1971}).

\bibitem[{\citenamefont{Kosterlitz and Thouless}(1973)}]{Kosterlitz1973}
\bibinfo{author}{\bibfnamefont{J.~M.} \bibnamefont{Kosterlitz}}
  \bibnamefont{and} \bibinfo{author}{\bibfnamefont{D.~J.}
  \bibnamefont{Thouless}}, \bibinfo{journal}{J. Phys. C}
  \textbf{\bibinfo{volume}{6}}, \bibinfo{pages}{1181} (\bibinfo{year}{1973}).

\bibitem[{\citenamefont{Kosterlitz}(1974)}]{Kosterlitz1974}
\bibinfo{author}{\bibfnamefont{J.~M.} \bibnamefont{Kosterlitz}},
  \bibinfo{journal}{J. Phys. C} \textbf{\bibinfo{volume}{7}},
  \bibinfo{pages}{1046} (\bibinfo{year}{1974}).

\bibitem[{\citenamefont{Hadzibabic et~al.}(2006)\citenamefont{Hadzibabic,
  Kr{\"u}ger, Cheneau, Battelier, and Dalibard}}]{DalibardBKT}
\bibinfo{author}{\bibfnamefont{Z.}~\bibnamefont{Hadzibabic}},
  \bibinfo{author}{\bibfnamefont{P.}~\bibnamefont{Kr{\"u}ger}},
  \bibinfo{author}{\bibfnamefont{M.}~\bibnamefont{Cheneau}},
  \bibinfo{author}{\bibfnamefont{B.}~\bibnamefont{Battelier}},
  \bibnamefont{and} \bibinfo{author}{\bibfnamefont{J.}~\bibnamefont{Dalibard}},
  \bibinfo{journal}{Nature} \textbf{\bibinfo{volume}{441}},
  \bibinfo{pages}{1118} (\bibinfo{year}{2006}).

\bibitem[{\citenamefont{Hebard and Fiory}(1980)}]{Hebard1980}
\bibinfo{author}{\bibfnamefont{A.}~\bibnamefont{Hebard}} \bibnamefont{and}
  \bibinfo{author}{\bibfnamefont{A.}~\bibnamefont{Fiory}},
  \bibinfo{journal}{Physical Review Letters} \textbf{\bibinfo{volume}{44}},
  \bibinfo{pages}{291} (\bibinfo{year}{1980}).

\bibitem[{\citenamefont{Halperin and Nelson}(1978)}]{Halperin1978}
\bibinfo{author}{\bibfnamefont{B.}~\bibnamefont{Halperin}} \bibnamefont{and}
  \bibinfo{author}{\bibfnamefont{D.~R.} \bibnamefont{Nelson}},
  \bibinfo{journal}{Physical Review Letters} \textbf{\bibinfo{volume}{41}},
  \bibinfo{pages}{121} (\bibinfo{year}{1978}).

\bibitem[{\citenamefont{Singh}(2000)}]{SinghRev}
\bibinfo{author}{\bibfnamefont{S.}~\bibnamefont{Singh}},
  \bibinfo{journal}{Physics Reports} \textbf{\bibinfo{volume}{324}},
  \bibinfo{pages}{107} (\bibinfo{year}{2000}).

\bibitem[{\citenamefont{Marchetti et~al.}(2013)\citenamefont{Marchetti, Joanny,
  Ramaswamy, Liverpool, Prost, Rao, and Simha}}]{MarchettiRev}
\bibinfo{author}{\bibfnamefont{M.~C.} \bibnamefont{Marchetti}},
  \bibinfo{author}{\bibfnamefont{J.-F.} \bibnamefont{Joanny}},
  \bibinfo{author}{\bibfnamefont{S.}~\bibnamefont{Ramaswamy}},
  \bibinfo{author}{\bibfnamefont{T.~B.} \bibnamefont{Liverpool}},
  \bibinfo{author}{\bibfnamefont{J.}~\bibnamefont{Prost}},
  \bibinfo{author}{\bibfnamefont{M.}~\bibnamefont{Rao}}, \bibnamefont{and}
  \bibinfo{author}{\bibfnamefont{R.~A.} \bibnamefont{Simha}},
  \bibinfo{journal}{Rev. Mod. Phys.} \textbf{\bibinfo{volume}{85}},
  \bibinfo{pages}{1143} (\bibinfo{year}{2013}).

\bibitem[{\citenamefont{Ramaswamy}(2010)}]{RamaswamyRev}
\bibinfo{author}{\bibfnamefont{S.}~\bibnamefont{Ramaswamy}},
  \bibinfo{journal}{Annu. Rev. Condens. Matter Phys.}
  \textbf{\bibinfo{volume}{1}}, \bibinfo{pages}{323} (\bibinfo{year}{2010}).

\bibitem[{\citenamefont{Doostmohammadi
  et~al.}(2018)\citenamefont{Doostmohammadi, Ign{\'e}s-Mullol, Yeomans, and
  Sagu{\'e}s}}]{SaguesRev}
\bibinfo{author}{\bibfnamefont{A.}~\bibnamefont{Doostmohammadi}},
  \bibinfo{author}{\bibfnamefont{J.}~\bibnamefont{Ign{\'e}s-Mullol}},
  \bibinfo{author}{\bibfnamefont{J.~M.} \bibnamefont{Yeomans}},
  \bibnamefont{and}
  \bibinfo{author}{\bibfnamefont{F.}~\bibnamefont{Sagu{\'e}s}},
  \bibinfo{journal}{Nature communications} \textbf{\bibinfo{volume}{9}},
  \bibinfo{pages}{1} (\bibinfo{year}{2018}).

\bibitem[{\citenamefont{Zhang et~al.}(2021{\natexlab{a}})\citenamefont{Zhang,
  Mozaffari, and de~Pablo}}]{ZhangRev}
\bibinfo{author}{\bibfnamefont{R.}~\bibnamefont{Zhang}},
  \bibinfo{author}{\bibfnamefont{A.}~\bibnamefont{Mozaffari}},
  \bibnamefont{and} \bibinfo{author}{\bibfnamefont{J.~J.}
  \bibnamefont{de~Pablo}}, \bibinfo{journal}{Nature Reviews Materials} pp.
  \bibinfo{pages}{1--17} (\bibinfo{year}{2021}{\natexlab{a}}).

\bibitem[{\citenamefont{Thampi et~al.}(2013)\citenamefont{Thampi, Golestanian,
  and Yeomans}}]{Thampi2013}
\bibinfo{author}{\bibfnamefont{S.~P.} \bibnamefont{Thampi}},
  \bibinfo{author}{\bibfnamefont{R.}~\bibnamefont{Golestanian}},
  \bibnamefont{and} \bibinfo{author}{\bibfnamefont{J.~M.}
  \bibnamefont{Yeomans}}, \bibinfo{journal}{Physical review letters}
  \textbf{\bibinfo{volume}{111}}, \bibinfo{pages}{118101}
  (\bibinfo{year}{2013}).

\bibitem[{\citenamefont{Giomi et~al.}(2013)\citenamefont{Giomi, Bowick, Ma, and
  Marchetti}}]{Giomi2013}
\bibinfo{author}{\bibfnamefont{L.}~\bibnamefont{Giomi}},
  \bibinfo{author}{\bibfnamefont{M.~J.} \bibnamefont{Bowick}},
  \bibinfo{author}{\bibfnamefont{X.}~\bibnamefont{Ma}}, \bibnamefont{and}
  \bibinfo{author}{\bibfnamefont{M.~C.} \bibnamefont{Marchetti}},
  \bibinfo{journal}{Physical review letters} \textbf{\bibinfo{volume}{110}},
  \bibinfo{pages}{228101} (\bibinfo{year}{2013}).

\bibitem[{\citenamefont{Giomi et~al.}(2014)\citenamefont{Giomi, Bowick, Mishra,
  Sknepnek, and Cristina~Marchetti}}]{Giomi2014}
\bibinfo{author}{\bibfnamefont{L.}~\bibnamefont{Giomi}},
  \bibinfo{author}{\bibfnamefont{M.~J.} \bibnamefont{Bowick}},
  \bibinfo{author}{\bibfnamefont{P.}~\bibnamefont{Mishra}},
  \bibinfo{author}{\bibfnamefont{R.}~\bibnamefont{Sknepnek}}, \bibnamefont{and}
  \bibinfo{author}{\bibfnamefont{M.}~\bibnamefont{Cristina~Marchetti}},
  \bibinfo{journal}{Philosophical Transactions of the Royal Society A:
  Mathematical, Physical and Engineering Sciences}
  \textbf{\bibinfo{volume}{372}}, \bibinfo{pages}{20130365}
  (\bibinfo{year}{2014}).

\bibitem[{\citenamefont{DeCamp et~al.}(2015)\citenamefont{DeCamp, Redner,
  Baskaran, Hagan, and Dogic}}]{Decamp2015}
\bibinfo{author}{\bibfnamefont{S.~J.} \bibnamefont{DeCamp}},
  \bibinfo{author}{\bibfnamefont{G.~S.} \bibnamefont{Redner}},
  \bibinfo{author}{\bibfnamefont{A.}~\bibnamefont{Baskaran}},
  \bibinfo{author}{\bibfnamefont{M.~F.} \bibnamefont{Hagan}}, \bibnamefont{and}
  \bibinfo{author}{\bibfnamefont{Z.}~\bibnamefont{Dogic}},
  \bibinfo{journal}{Nature materials} \textbf{\bibinfo{volume}{14}},
  \bibinfo{pages}{1110} (\bibinfo{year}{2015}).

\bibitem[{\citenamefont{Li et~al.}(2019)\citenamefont{Li, Shi, Huang, Chen,
  Xiao, Liu, Chat{\'e}, and Zhang}}]{ChateBacteria2019}
\bibinfo{author}{\bibfnamefont{H.}~\bibnamefont{Li}},
  \bibinfo{author}{\bibfnamefont{X.-q.} \bibnamefont{Shi}},
  \bibinfo{author}{\bibfnamefont{M.}~\bibnamefont{Huang}},
  \bibinfo{author}{\bibfnamefont{X.}~\bibnamefont{Chen}},
  \bibinfo{author}{\bibfnamefont{M.}~\bibnamefont{Xiao}},
  \bibinfo{author}{\bibfnamefont{C.}~\bibnamefont{Liu}},
  \bibinfo{author}{\bibfnamefont{H.}~\bibnamefont{Chat{\'e}}},
  \bibnamefont{and} \bibinfo{author}{\bibfnamefont{H.}~\bibnamefont{Zhang}},
  \bibinfo{journal}{Proceedings of the National Academy of Sciences}
  \textbf{\bibinfo{volume}{116}}, \bibinfo{pages}{777} (\bibinfo{year}{2019}).

\bibitem[{\citenamefont{Shankar et~al.}(2018)\citenamefont{Shankar, Ramaswamy,
  Marchetti, and Bowick}}]{Shankar2018}
\bibinfo{author}{\bibfnamefont{S.}~\bibnamefont{Shankar}},
  \bibinfo{author}{\bibfnamefont{S.}~\bibnamefont{Ramaswamy}},
  \bibinfo{author}{\bibfnamefont{M.~C.} \bibnamefont{Marchetti}},
  \bibnamefont{and} \bibinfo{author}{\bibfnamefont{M.~J.}
  \bibnamefont{Bowick}}, \bibinfo{journal}{Physical review letters}
  \textbf{\bibinfo{volume}{121}}, \bibinfo{pages}{108002}
  (\bibinfo{year}{2018}).

\bibitem[{\citenamefont{Shankar and Marchetti}(2019)}]{Shankar2019}
\bibinfo{author}{\bibfnamefont{S.}~\bibnamefont{Shankar}} \bibnamefont{and}
  \bibinfo{author}{\bibfnamefont{M.~C.} \bibnamefont{Marchetti}},
  \bibinfo{journal}{Physical Review X} \textbf{\bibinfo{volume}{9}},
  \bibinfo{pages}{041047} (\bibinfo{year}{2019}).

\bibitem[{\citenamefont{Vliegenthart et~al.}(2020)\citenamefont{Vliegenthart,
  Ravichandran, Ripoll, Auth, and Gompper}}]{Gompper2020}
\bibinfo{author}{\bibfnamefont{G.~A.} \bibnamefont{Vliegenthart}},
  \bibinfo{author}{\bibfnamefont{A.}~\bibnamefont{Ravichandran}},
  \bibinfo{author}{\bibfnamefont{M.}~\bibnamefont{Ripoll}},
  \bibinfo{author}{\bibfnamefont{T.}~\bibnamefont{Auth}}, \bibnamefont{and}
  \bibinfo{author}{\bibfnamefont{G.}~\bibnamefont{Gompper}},
  \bibinfo{journal}{Science advances} \textbf{\bibinfo{volume}{6}},
  \bibinfo{pages}{eaaw9975} (\bibinfo{year}{2020}).

\bibitem[{\citenamefont{Weber et~al.}(2014)\citenamefont{Weber, Bock, and
  Frey}}]{Weber2014}
\bibinfo{author}{\bibfnamefont{C.~A.} \bibnamefont{Weber}},
  \bibinfo{author}{\bibfnamefont{C.}~\bibnamefont{Bock}}, \bibnamefont{and}
  \bibinfo{author}{\bibfnamefont{E.}~\bibnamefont{Frey}},
  \bibinfo{journal}{Phys. Rev. Lett.} \textbf{\bibinfo{volume}{112}},
  \bibinfo{pages}{168301} (\bibinfo{year}{2014}).

\bibitem[{\citenamefont{Paliwal and Dijkstra}(2020)}]{PaliwalDijkstra}
\bibinfo{author}{\bibfnamefont{S.}~\bibnamefont{Paliwal}} \bibnamefont{and}
  \bibinfo{author}{\bibfnamefont{M.}~\bibnamefont{Dijkstra}},
  \bibinfo{journal}{Phys. Rev. Research} \textbf{\bibinfo{volume}{2}},
  \bibinfo{pages}{012013} (\bibinfo{year}{2020}).

\bibitem[{\citenamefont{Digregorio et~al.}(2019)\citenamefont{Digregorio,
  Levis, Cugliandolo, Gonnella, and Pagonabarraga}}]{LinoDefects}
\bibinfo{author}{\bibfnamefont{P.}~\bibnamefont{Digregorio}},
  \bibinfo{author}{\bibfnamefont{D.}~\bibnamefont{Levis}},
  \bibinfo{author}{\bibfnamefont{L.~F.} \bibnamefont{Cugliandolo}},
  \bibinfo{author}{\bibfnamefont{G.}~\bibnamefont{Gonnella}}, \bibnamefont{and}
  \bibinfo{author}{\bibfnamefont{I.}~\bibnamefont{Pagonabarraga}},
  \bibinfo{journal}{arXiv preprint arXiv:1911.06366}  (\bibinfo{year}{2019}).

\bibitem[{\citenamefont{Bililign et~al.}(2021)\citenamefont{Bililign, Usabiaga,
  Ganan, Soni, Magkiriadou, Shelley, Bartolo, and Irvine}}]{Bartolo2021}
\bibinfo{author}{\bibfnamefont{E.~S.} \bibnamefont{Bililign}},
  \bibinfo{author}{\bibfnamefont{F.~B.} \bibnamefont{Usabiaga}},
  \bibinfo{author}{\bibfnamefont{Y.~A.} \bibnamefont{Ganan}},
  \bibinfo{author}{\bibfnamefont{V.}~\bibnamefont{Soni}},
  \bibinfo{author}{\bibfnamefont{S.}~\bibnamefont{Magkiriadou}},
  \bibinfo{author}{\bibfnamefont{M.~J.} \bibnamefont{Shelley}},
  \bibinfo{author}{\bibfnamefont{D.}~\bibnamefont{Bartolo}}, \bibnamefont{and}
  \bibinfo{author}{\bibfnamefont{W.}~\bibnamefont{Irvine}},
  \bibinfo{journal}{arXiv preprint arXiv:2102.03263}  (\bibinfo{year}{2021}).

\bibitem[{\citenamefont{Chardac et~al.}(2021)\citenamefont{Chardac, Hoffmann,
  Poupart, Giomi, and Bartolo}}]{ChardacBartolo2021}
\bibinfo{author}{\bibfnamefont{A.}~\bibnamefont{Chardac}},
  \bibinfo{author}{\bibfnamefont{L.~A.} \bibnamefont{Hoffmann}},
  \bibinfo{author}{\bibfnamefont{Y.}~\bibnamefont{Poupart}},
  \bibinfo{author}{\bibfnamefont{L.}~\bibnamefont{Giomi}}, \bibnamefont{and}
  \bibinfo{author}{\bibfnamefont{D.}~\bibnamefont{Bartolo}},
  \bibinfo{journal}{arXiv preprint arXiv:2103.03861}  (\bibinfo{year}{2021}).

\bibitem[{\citenamefont{Kuramoto}(2012)}]{KuramotoBook}
\bibinfo{author}{\bibfnamefont{Y.}~\bibnamefont{Kuramoto}},
  \emph{\bibinfo{title}{Chemical oscillations, waves, and turbulence}},
  vol.~\bibinfo{volume}{19} (\bibinfo{publisher}{Springer Science \& Business
  Media}, \bibinfo{year}{2012}).

\bibitem[{\citenamefont{Acebr{\'o}n et~al.}(2005)\citenamefont{Acebr{\'o}n,
  Bonilla, Perez-Vicente, Ritort, and Spigler}}]{AcebronRev}
\bibinfo{author}{\bibfnamefont{J.~A.} \bibnamefont{Acebr{\'o}n}},
  \bibinfo{author}{\bibfnamefont{L.~L.} \bibnamefont{Bonilla}},
  \bibinfo{author}{\bibfnamefont{C.~J.} \bibnamefont{Perez-Vicente}},
  \bibinfo{author}{\bibfnamefont{F.}~\bibnamefont{Ritort}}, \bibnamefont{and}
  \bibinfo{author}{\bibfnamefont{R.}~\bibnamefont{Spigler}},
  \bibinfo{journal}{Rev. Mod. Phys.} \textbf{\bibinfo{volume}{77}},
  \bibinfo{pages}{137} (\bibinfo{year}{2005}).

\bibitem[{\citenamefont{Strogatz}(2004)}]{StrogatzBook}
\bibinfo{author}{\bibfnamefont{S.}~\bibnamefont{Strogatz}},
  \emph{\bibinfo{title}{Sync: The Emerging Science of Spontaneous Order}}
  (\bibinfo{publisher}{Penguin UK}, \bibinfo{year}{2004}).

\bibitem[{\citenamefont{Pikovsky et~al.}(2003)\citenamefont{Pikovsky,
  Rosenblum, and Kurths}}]{Pikovsky2003}
\bibinfo{author}{\bibfnamefont{A.}~\bibnamefont{Pikovsky}},
  \bibinfo{author}{\bibfnamefont{M.}~\bibnamefont{Rosenblum}},
  \bibnamefont{and} \bibinfo{author}{\bibfnamefont{J.}~\bibnamefont{Kurths}},
  \emph{\bibinfo{title}{Synchronization: a universal concept in nonlinear
  sciences}}, vol.~\bibinfo{volume}{12} (\bibinfo{publisher}{Cambridge
  University Press}, \bibinfo{year}{2003}).

\bibitem[{\citenamefont{Kuramoto}(1975)}]{KuramotoOriginal}
\bibinfo{author}{\bibfnamefont{Y.}~\bibnamefont{Kuramoto}}, in
  \emph{\bibinfo{booktitle}{International symposium on mathematical problems in
  theoretical physics}} (\bibinfo{organization}{Springer},
  \bibinfo{year}{1975}), pp. \bibinfo{pages}{420--422}.

\bibitem[{\citenamefont{Arenas et~al.}(2008)\citenamefont{Arenas,
  D\'{\i}az-Guilera, Kurths, Moreno, and Zhou}}]{ArenasRev}
\bibinfo{author}{\bibfnamefont{A.}~\bibnamefont{Arenas}},
  \bibinfo{author}{\bibfnamefont{A.}~\bibnamefont{D\'{\i}az-Guilera}},
  \bibinfo{author}{\bibfnamefont{J.}~\bibnamefont{Kurths}},
  \bibinfo{author}{\bibfnamefont{Y.}~\bibnamefont{Moreno}}, \bibnamefont{and}
  \bibinfo{author}{\bibfnamefont{C.}~\bibnamefont{Zhou}},
  \bibinfo{journal}{Phys. Rep.} \textbf{\bibinfo{volume}{469}},
  \bibinfo{pages}{93} (\bibinfo{year}{2008}).

\bibitem[{\citenamefont{Gupta et~al.}(2014)\citenamefont{Gupta, Campa, and
  Ruffo}}]{RuffoRev}
\bibinfo{author}{\bibfnamefont{S.}~\bibnamefont{Gupta}},
  \bibinfo{author}{\bibfnamefont{A.}~\bibnamefont{Campa}}, \bibnamefont{and}
  \bibinfo{author}{\bibfnamefont{S.}~\bibnamefont{Ruffo}},
  \bibinfo{journal}{Journal of Statistical Mechanics: Theory and Experiment}
  \textbf{\bibinfo{volume}{2014}}, \bibinfo{pages}{R08001}
  (\bibinfo{year}{2014}).

\bibitem[{\citenamefont{Strogatz and Mirollo}(1988)}]{Strogatz1988}
\bibinfo{author}{\bibfnamefont{S.~H.} \bibnamefont{Strogatz}} \bibnamefont{and}
  \bibinfo{author}{\bibfnamefont{R.~E.} \bibnamefont{Mirollo}},
  \bibinfo{journal}{J. Phys. A} \textbf{\bibinfo{volume}{21}},
  \bibinfo{pages}{L699} (\bibinfo{year}{1988}).

\bibitem[{\citenamefont{Sakaguchi et~al.}(1987)\citenamefont{Sakaguchi,
  Shinomoto, and Kuramoto}}]{Sakaguchi1987}
\bibinfo{author}{\bibfnamefont{H.}~\bibnamefont{Sakaguchi}},
  \bibinfo{author}{\bibfnamefont{S.}~\bibnamefont{Shinomoto}},
  \bibnamefont{and} \bibinfo{author}{\bibfnamefont{Y.}~\bibnamefont{Kuramoto}},
  \bibinfo{journal}{Prog. Theor. Phys.} \textbf{\bibinfo{volume}{77}},
  \bibinfo{pages}{1005} (\bibinfo{year}{1987}).

\bibitem[{\citenamefont{Daido}(1988)}]{Daido1988}
\bibinfo{author}{\bibfnamefont{H.}~\bibnamefont{Daido}},
  \bibinfo{journal}{Phys. Rev. Lett.} \textbf{\bibinfo{volume}{61}},
  \bibinfo{pages}{231} (\bibinfo{year}{1988}).

\bibitem[{\citenamefont{Niebur et~al.}(1991)\citenamefont{Niebur, Schuster,
  Kammen, and Koch}}]{Niebur1991}
\bibinfo{author}{\bibfnamefont{E.}~\bibnamefont{Niebur}},
  \bibinfo{author}{\bibfnamefont{H.~G.} \bibnamefont{Schuster}},
  \bibinfo{author}{\bibfnamefont{D.~M.} \bibnamefont{Kammen}},
  \bibnamefont{and} \bibinfo{author}{\bibfnamefont{C.}~\bibnamefont{Koch}},
  \bibinfo{journal}{Phys. Rev. A} \textbf{\bibinfo{volume}{44}},
  \bibinfo{pages}{6895} (\bibinfo{year}{1991}).

\bibitem[{\citenamefont{Lee et~al.}(2010)\citenamefont{Lee, Tam, Refael,
  Rogers, and Cross.}}]{Lee}
\bibinfo{author}{\bibnamefont{Lee}}, \bibinfo{author}{\bibnamefont{Tam}},
  \bibinfo{author}{\bibnamefont{Refael}},
  \bibinfo{author}{\bibnamefont{Rogers}}, \bibnamefont{and}
  \bibinfo{author}{\bibnamefont{Cross.}}, \bibinfo{journal}{Physical Review E}
  \textbf{\bibinfo{volume}{82}} (\bibinfo{year}{2010}).

\bibitem[{\citenamefont{Hong et~al.}(2005)\citenamefont{Hong, Park, and
  Choi}}]{Hong2005}
\bibinfo{author}{\bibfnamefont{H.}~\bibnamefont{Hong}},
  \bibinfo{author}{\bibfnamefont{H.}~\bibnamefont{Park}}, \bibnamefont{and}
  \bibinfo{author}{\bibfnamefont{M.~Y.} \bibnamefont{Choi}},
  \bibinfo{journal}{Phys. Rev. E} \textbf{\bibinfo{volume}{72}},
  \bibinfo{pages}{036217} (\bibinfo{year}{2005}).

\bibitem[{\citenamefont{Hong et~al.}(2007)\citenamefont{Hong, Chat{\'e}, Park,
  and Tang}}]{Hong2007}
\bibinfo{author}{\bibfnamefont{H.}~\bibnamefont{Hong}},
  \bibinfo{author}{\bibfnamefont{H.}~\bibnamefont{Chat{\'e}}},
  \bibinfo{author}{\bibfnamefont{H.}~\bibnamefont{Park}}, \bibnamefont{and}
  \bibinfo{author}{\bibfnamefont{L.-H.} \bibnamefont{Tang}},
  \bibinfo{journal}{Phys. Rev. Lett.} \textbf{\bibinfo{volume}{99}},
  \bibinfo{pages}{184101} (\bibinfo{year}{2007}).

\bibitem[{\citenamefont{Sarkar and Gupte}(2021)}]{Sarkar2021}
\bibinfo{author}{\bibfnamefont{M.}~\bibnamefont{Sarkar}} \bibnamefont{and}
  \bibinfo{author}{\bibfnamefont{N.}~\bibnamefont{Gupte}},
  \bibinfo{journal}{Physical Review E} \textbf{\bibinfo{volume}{103}},
  \bibinfo{pages}{032204} (\bibinfo{year}{2021}).

\bibitem[{SM()}]{SM}
\bibinfo{journal}{See Supplemental Material at doi:...}  (????).

\bibitem[{\citenamefont{Loft and DeGrand}(1987)}]{LoftDeGrand}
\bibinfo{author}{\bibfnamefont{R.}~\bibnamefont{Loft}} \bibnamefont{and}
  \bibinfo{author}{\bibfnamefont{T.~A.} \bibnamefont{DeGrand}},
  \bibinfo{journal}{Physical Review B} \textbf{\bibinfo{volume}{35}},
  \bibinfo{pages}{8528} (\bibinfo{year}{1987}).

\bibitem[{\citenamefont{Yurke et~al.}(1993)\citenamefont{Yurke, Pargellis,
  Kovacs, and Huse}}]{YurkeHuse1993}
\bibinfo{author}{\bibfnamefont{B.}~\bibnamefont{Yurke}},
  \bibinfo{author}{\bibfnamefont{A.}~\bibnamefont{Pargellis}},
  \bibinfo{author}{\bibfnamefont{T.}~\bibnamefont{Kovacs}}, \bibnamefont{and}
  \bibinfo{author}{\bibfnamefont{D.}~\bibnamefont{Huse}},
  \bibinfo{journal}{Physical Review E} \textbf{\bibinfo{volume}{47}},
  \bibinfo{pages}{1525} (\bibinfo{year}{1993}).

\bibitem[{\citenamefont{Paoluzzi et~al.}(2018)\citenamefont{Paoluzzi, Marconi,
  and Maggi}}]{Paoluzzi2018}
\bibinfo{author}{\bibfnamefont{M.}~\bibnamefont{Paoluzzi}},
  \bibinfo{author}{\bibfnamefont{U.~M.~B.} \bibnamefont{Marconi}},
  \bibnamefont{and} \bibinfo{author}{\bibfnamefont{C.}~\bibnamefont{Maggi}},
  \bibinfo{journal}{Physical Review E} \textbf{\bibinfo{volume}{97}},
  \bibinfo{pages}{022605} (\bibinfo{year}{2018}).

\bibitem[{\citenamefont{Gupta et~al.}(1988)\citenamefont{Gupta, DeLapp,
  Batrouni, Fox, Baillie, and Apostolakis}}]{Gupta1988}
\bibinfo{author}{\bibfnamefont{R.}~\bibnamefont{Gupta}},
  \bibinfo{author}{\bibfnamefont{J.}~\bibnamefont{DeLapp}},
  \bibinfo{author}{\bibfnamefont{G.~G.} \bibnamefont{Batrouni}},
  \bibinfo{author}{\bibfnamefont{G.~C.} \bibnamefont{Fox}},
  \bibinfo{author}{\bibfnamefont{C.~F.} \bibnamefont{Baillie}},
  \bibnamefont{and}
  \bibinfo{author}{\bibfnamefont{J.}~\bibnamefont{Apostolakis}},
  \bibinfo{journal}{Physical review letters} \textbf{\bibinfo{volume}{61}},
  \bibinfo{pages}{1996} (\bibinfo{year}{1988}).

\bibitem[{\citenamefont{Hasenbusch}(2005)}]{Hasenbusch2005}
\bibinfo{author}{\bibfnamefont{M.}~\bibnamefont{Hasenbusch}},
  \bibinfo{journal}{Journal of Physics A: Mathematical and General}
  \textbf{\bibinfo{volume}{38}}, \bibinfo{pages}{5869} (\bibinfo{year}{2005}).

\bibitem[{\citenamefont{Kuhn}(1955)}]{Hungarian}
\bibinfo{author}{\bibnamefont{Kuhn}}, \bibinfo{journal}{Naval Research
  Logistics Quarterly}  (\bibinfo{year}{1955}).

\bibitem[{\citenamefont{Munkres}(1957)}]{HungarianComplexity}
\bibinfo{author}{\bibnamefont{Munkres}}, \bibinfo{journal}{Journal of the
  Society for Industrial and Applied Mathematics}  (\bibinfo{year}{1957}).

\bibitem[{\citenamefont{Tomizawa}(1971)}]{HungarianImproved}
\bibinfo{author}{\bibnamefont{Tomizawa}}, \bibinfo{journal}{Networks}
  (\bibinfo{year}{1971}).

\bibitem[{\citenamefont{Olsson}(1992)}]{Olsson1992}
\bibinfo{author}{\bibfnamefont{P.}~\bibnamefont{Olsson}},
  \bibinfo{journal}{Physical Review B} \textbf{\bibinfo{volume}{46}},
  \bibinfo{pages}{14598} (\bibinfo{year}{1992}).

\bibitem[{\citenamefont{Kim et~al.}(1999)\citenamefont{Kim, Minnhagen, and
  Olsson}}]{Olsson1999}
\bibinfo{author}{\bibfnamefont{B.~J.} \bibnamefont{Kim}},
  \bibinfo{author}{\bibfnamefont{P.}~\bibnamefont{Minnhagen}},
  \bibnamefont{and} \bibinfo{author}{\bibfnamefont{P.}~\bibnamefont{Olsson}},
  \bibinfo{journal}{Physical Review B} \textbf{\bibinfo{volume}{59}},
  \bibinfo{pages}{11506} (\bibinfo{year}{1999}).

\bibitem[{\citenamefont{Nogawa and Nemoto}(2009)}]{Japo}
\bibinfo{author}{\bibfnamefont{T.}~\bibnamefont{Nogawa}} \bibnamefont{and}
  \bibinfo{author}{\bibfnamefont{K.}~\bibnamefont{Nemoto}},
  \bibinfo{journal}{Journal of the Physical Society of Japan}
  \textbf{\bibinfo{volume}{78}}, \bibinfo{pages}{064001}
  (\bibinfo{year}{2009}).

\bibitem[{\citenamefont{Redner}(1989)}]{Redner1989superdiff}
\bibinfo{author}{\bibfnamefont{S.}~\bibnamefont{Redner}},
  \bibinfo{journal}{Physica D: Nonlinear Phenomena}
  \textbf{\bibinfo{volume}{38}}, \bibinfo{pages}{287} (\bibinfo{year}{1989}).

\bibitem[{\citenamefont{Redner}(1990)}]{Redner1990superdiff}
\bibinfo{author}{\bibfnamefont{S.}~\bibnamefont{Redner}},
  \bibinfo{journal}{Physica A: Statistical Mechanics and its Applications}
  \textbf{\bibinfo{volume}{168}}, \bibinfo{pages}{551} (\bibinfo{year}{1990}).

\bibitem[{\citenamefont{Bouchaud et~al.}(1990)\citenamefont{Bouchaud, Georges,
  Koplik, Provata, and Redner}}]{BouchaudGeorgesPRL1990}
\bibinfo{author}{\bibfnamefont{J.-P.} \bibnamefont{Bouchaud}},
  \bibinfo{author}{\bibfnamefont{A.}~\bibnamefont{Georges}},
  \bibinfo{author}{\bibfnamefont{J.}~\bibnamefont{Koplik}},
  \bibinfo{author}{\bibfnamefont{A.}~\bibnamefont{Provata}}, \bibnamefont{and}
  \bibinfo{author}{\bibfnamefont{S.}~\bibnamefont{Redner}},
  \bibinfo{journal}{Physical review letters} \textbf{\bibinfo{volume}{64}},
  \bibinfo{pages}{2503} (\bibinfo{year}{1990}).

\bibitem[{\citenamefont{Nezhadhaghighi}(2020)}]{PhysA2020}
\bibinfo{author}{\bibfnamefont{M.~G.} \bibnamefont{Nezhadhaghighi}},
  \bibinfo{journal}{Physica A: Statistical Mechanics and its Applications}
  \textbf{\bibinfo{volume}{557}}, \bibinfo{pages}{124977}
  (\bibinfo{year}{2020}).

\bibitem[{\citenamefont{Bouchaud and Georges}(1990)}]{BouchaudRev}
\bibinfo{author}{\bibfnamefont{J.-P.} \bibnamefont{Bouchaud}} \bibnamefont{and}
  \bibinfo{author}{\bibfnamefont{A.}~\bibnamefont{Georges}},
  \bibinfo{journal}{Physics reports} \textbf{\bibinfo{volume}{195}},
  \bibinfo{pages}{127} (\bibinfo{year}{1990}).

\bibitem[{\citenamefont{You et~al.}(2018)\citenamefont{You, Pearce, Sengupta,
  and Giomi}}]{GiomiPierce}
\bibinfo{author}{\bibfnamefont{Z.}~\bibnamefont{You}},
  \bibinfo{author}{\bibfnamefont{D.~J.} \bibnamefont{Pearce}},
  \bibinfo{author}{\bibfnamefont{A.}~\bibnamefont{Sengupta}}, \bibnamefont{and}
  \bibinfo{author}{\bibfnamefont{L.}~\bibnamefont{Giomi}},
  \bibinfo{journal}{Phys. Rev. X} \textbf{\bibinfo{volume}{8}},
  \bibinfo{pages}{031065} (\bibinfo{year}{2018}).

\bibitem[{\citenamefont{Caporusso et~al.}(2020)\citenamefont{Caporusso,
  Digregorio, Levis, Cugliandolo, and Gonnella}}]{ClaudioPRL}
\bibinfo{author}{\bibfnamefont{C.~B.} \bibnamefont{Caporusso}},
  \bibinfo{author}{\bibfnamefont{P.}~\bibnamefont{Digregorio}},
  \bibinfo{author}{\bibfnamefont{D.}~\bibnamefont{Levis}},
  \bibinfo{author}{\bibfnamefont{L.~F.} \bibnamefont{Cugliandolo}},
  \bibnamefont{and} \bibinfo{author}{\bibfnamefont{G.}~\bibnamefont{Gonnella}},
  \bibinfo{journal}{Physical Review Letters} \textbf{\bibinfo{volume}{125}},
  \bibinfo{pages}{178004} (\bibinfo{year}{2020}).

\bibitem[{\citenamefont{Bruot and Cicuta}(2016)}]{CicutaSyncRev}
\bibinfo{author}{\bibfnamefont{N.}~\bibnamefont{Bruot}} \bibnamefont{and}
  \bibinfo{author}{\bibfnamefont{P.}~\bibnamefont{Cicuta}},
  \bibinfo{journal}{Annual Review of Condensed Matter Physics}
  \textbf{\bibinfo{volume}{7}}, \bibinfo{pages}{323} (\bibinfo{year}{2016}).

\bibitem[{\citenamefont{Zhang et~al.}(2021{\natexlab{b}})\citenamefont{Zhang,
  Yuan, Dou, de~la Cruz, and Bishop}}]{zhang2021quincke}
\bibinfo{author}{\bibfnamefont{Z.}~\bibnamefont{Zhang}},
  \bibinfo{author}{\bibfnamefont{H.}~\bibnamefont{Yuan}},
  \bibinfo{author}{\bibfnamefont{Y.}~\bibnamefont{Dou}},
  \bibinfo{author}{\bibfnamefont{M.~O.} \bibnamefont{de~la Cruz}},
  \bibnamefont{and} \bibinfo{author}{\bibfnamefont{K.~J.}
  \bibnamefont{Bishop}}, \bibinfo{journal}{arXiv preprint arXiv:2102.05237}
  (\bibinfo{year}{2021}{\natexlab{b}}).

\bibitem[{\citenamefont{Mondrag{\'o}n-Palomino
  et~al.}(2011)\citenamefont{Mondrag{\'o}n-Palomino, Danino, Selimkhanov,
  Tsimring, and Hasty}}]{Mondragon2011}
\bibinfo{author}{\bibfnamefont{O.}~\bibnamefont{Mondrag{\'o}n-Palomino}},
  \bibinfo{author}{\bibfnamefont{T.}~\bibnamefont{Danino}},
  \bibinfo{author}{\bibfnamefont{J.}~\bibnamefont{Selimkhanov}},
  \bibinfo{author}{\bibfnamefont{L.}~\bibnamefont{Tsimring}}, \bibnamefont{and}
  \bibinfo{author}{\bibfnamefont{J.}~\bibnamefont{Hasty}},
  \bibinfo{journal}{Science} \textbf{\bibinfo{volume}{333}},
  \bibinfo{pages}{1315} (\bibinfo{year}{2011}).

\bibitem[{\citenamefont{Prindle et~al.}(2014)\citenamefont{Prindle,
  Selimkhanov, Li, Razinkov, Tsimring, and Hasty}}]{Prindle2014}
\bibinfo{author}{\bibfnamefont{A.}~\bibnamefont{Prindle}},
  \bibinfo{author}{\bibfnamefont{J.}~\bibnamefont{Selimkhanov}},
  \bibinfo{author}{\bibfnamefont{H.}~\bibnamefont{Li}},
  \bibinfo{author}{\bibfnamefont{I.}~\bibnamefont{Razinkov}},
  \bibinfo{author}{\bibfnamefont{L.~S.} \bibnamefont{Tsimring}},
  \bibnamefont{and} \bibinfo{author}{\bibfnamefont{J.}~\bibnamefont{Hasty}},
  \bibinfo{journal}{Nature} \textbf{\bibinfo{volume}{508}},
  \bibinfo{pages}{387} (\bibinfo{year}{2014}).

\bibitem[{\citenamefont{Flovik et~al.}(2016)\citenamefont{Flovik, Macia, and
  Wahlstr{\"o}m}}]{Ferran2016}
\bibinfo{author}{\bibfnamefont{V.}~\bibnamefont{Flovik}},
  \bibinfo{author}{\bibfnamefont{F.}~\bibnamefont{Macia}}, \bibnamefont{and}
  \bibinfo{author}{\bibfnamefont{E.}~\bibnamefont{Wahlstr{\"o}m}},
  \bibinfo{journal}{Scientific reports} \textbf{\bibinfo{volume}{6}},
  \bibinfo{pages}{1} (\bibinfo{year}{2016}).

\end{thebibliography}

\end{document}